\documentstyle[11pt,aaspp]{article}

\tighten

\input psfig.tex

\def\check_mode#1{\ifmmode{#1}\else{$#1$}\fi}

\def\ddeg   {\check_mode{{\rlap.}^\circ}}
\def\lsim   {\check_mode{_<\atop^{\sim}}}

\slugcomment{COBE Preprint \# 96-06, Submitted to {\it ApJ Letters}}

\begin{document}

\title{2-Point Correlations in the {\it COBE}\altaffilmark{1} 
DMR\\4-Year Anisotropy Maps}

\author{ G. Hinshaw\altaffilmark{2,3},
         A.J. Banday\altaffilmark{2,4},
         C.L. Bennett\altaffilmark{5},
         K.M. G\'orski\altaffilmark{2,6},
         A. Kogut\altaffilmark{2}, \nl
         C.H. Lineweaver\altaffilmark{7},
         G.F. Smoot\altaffilmark{8,9}
      \& E.L. Wright\altaffilmark{10}}

\altaffiltext{1}{The National Aeronautics and Space Administration/Goddard 
Space Flight Center (NASA/GSFC) is responsible for the design, development, 
and operation of the Cosmic Background Explorer ({\it COBE}).
Scientific guidance is provided by the {\it COBE} Science Working Group.
GSFC is also responsible for the development of the analysis software and
for the production of the mission data sets.}
\altaffiltext{2}{Hughes STX Corporation, 
                 Laboratory for Astronomy and Solar Physics,
                 Code 685, NASA/GSFC, 
                 Greenbelt MD 20771.}
\altaffiltext{3}{e-mail: hinshaw@stars.gsfc.nasa.gov}
\altaffiltext{4}{Current address:
                 Max Planck Institut fur Astrophysik,
                 85740 Garching Bei Munchen,
                 Germany.}
\altaffiltext{5}{Laboratory for Astronomy and Solar Physics,
                 Code 685, NASA/GSFC, 
                 Greenbelt MD 20771.}
\altaffiltext{6}{On leave from Warsaw University Observatory,
                 Aleje Ujazdowskie 4, 00-478 Warszawa, Poland.}
\altaffiltext{7}{Observatoire de Strasbourg, 67000, Strasbourg, France.}
\altaffiltext{8}{Department of Physics, U.C. Berkeley, Berkeley CA 94720.}
\altaffiltext{9}{Lawrence Berkeley Laboratory, Bldg 50-351, 
                 University of California, Berkeley CA 94720.}
\altaffiltext{10}{UCLA Astronomy, P.O. Box 951562, Los Angeles CA 90095-1562.}

\begin{abstract}
The 2-point temperature correlation function is evaluated from 
the 4-year {\tenit COBE} DMR microwave anisotropy maps.  We examine the 
2-point function, which is the Legendre transform of the angular power 
spectrum, and show that the data are statistically consistent from channel 
to channel and frequency to frequency.  The most likely quadrupole 
normalization is computed for a scale-invariant power-law spectrum
of CMB anisotropy, using a variety of data combinations.  For a given data set, 
the normalization inferred from the 2-point data is consistent with that 
inferred by other methods.  The smallest and largest normalization deduced 
from any data combination are 16.4 and 19.6 $\mu$K respectively, with a value
$\sim$18 $\mu$K generally preferred.
\end{abstract}


\clearpage
\section{Introduction}
The detection of large angular scale anisotropies in the Cosmic Microwave 
Background (CMB) radiation was first reported by the {\it COBE}-DMR experiment
in 1992 (Smoot et al. 1992; Bennett et al. 1992; Wright et al. 1992; Kogut et 
al. 1992).  The initial detection was based only on the first year of flight 
data.  Since that time the DMR Team processed and analyzed the first two 
years of data and found results to be consistent with the first year 
results (Bennett et al. 1994, G\'orski et al. 1994, Wright et al. 1994).  We 
have now processed and analyzed the full 4-years of DMR observations: this 
paper is one of a series describing the results of our analysis.  The maps and 
an overview of the scientific results are given in Bennett et al. (1996).

In this paper we analyze the anisotropy in the 4-year DMR maps using the 
2-point correlation function as a measure of the angular power spectrum.  
The {\it COBE}-DMR experiment was designed to measure the CMB anisotropy on 
angular scales of $\geq 7^{\circ}$, corresponding to spherical harmonic 
multipole moments of order $\ell\,\lsim\,30$.  The DMR has produced full-sky 
maps of the CMB temperature at each of three frequencies, 31.5, 53, and 90 
GHz, with two independent channels, A and B, at each frequency.  In principal, 
one can obtain an estimate of the CMB power spectrum from an anisotropy map 
simply by decomposing the map, $T(\theta,\phi)$, into spherical harmonic 
components and averaging them to find the mean power per mode $\ell$:
$T(\theta,\phi) = \sum_{\ell,m} a_{\ell m} Y_{\ell m}(\theta,\phi)$ with power 
spectrum $a_{\ell}^2 = \frac{1}{2\ell+1} \sum_{m=-\ell}^{\ell} |a_{\ell m}|^2$.
In practice, however, there are a number of complications that arise.  First, 
the need to apply a galactic cut to the data renders the spherical harmonics 
non-orthogonal, thereby coupling the $a_{\ell m}$ coefficients and increasing 
their uncertainty.  Moreover, since $a_{\ell}^2$ is a quadratic form, any 
uncertainty in the $a_{\ell m}$ (whether due to coupling, instrument noise, 
systematic effects and/or foreground sources) produces a positive bias 
in the estimate of the power spectrum.  However, see G\'orski et al. (1996) 
and Wright et al. (1996) for spherical harmonic-based analyses that account 
for these difficulties.

An alternative to estimating the power spectrum is to evaluate its Legendre 
transform, the 2-point correlation function.  For a given power spectrum with 
multipole amplitudes $C_{\ell} = \langle |a_{\ell m}|^2 \rangle$, the predicted 
covariance between pairs of map pixels $i$ and $j$ with angular separation 
$\alpha_{ij}$ is
\begin{equation}
C(\alpha_{ij}) = \langle T_i T_j \rangle = \frac{1}{4\pi} \sum_{\ell} \, 
(2\ell+1) \, W^{2}_{\ell} \, C_{\ell} \, P_l(\cos \alpha_{ij})
\end{equation}
where $T_i$ is the CMB temperature in pixel $i$, the angled brackets denote an 
average over an ensemble of universal observers, $W_{\ell}$ is the 
experimental window function that includes the effects of beam smoothing and 
finite pixel size, and $P_l(\cos \alpha_{ij})$ is the Legendre polynomial 
of order $\ell$.  We estimate the 2-point correlation function in our sky 
by evaluating the average product of all map temperatures with a fixed angular 
separation
\begin{equation}
C(\alpha) = \sum_{i,j} w_iw_j\ T_iT_j ~/~ \sum_{i,j} w_iw_j.
\label{2pt_data_eq}
\end{equation}
where the sum is restricted to pixel pairs ($i,j$) separated by an angle 
$\alpha$, $T_i$ is the observed temperature in pixel $i$ after monopole and 
dipole (and optionally quadrupole) subtraction, and $w_i$ is the statistical 
weight of pixel $i$. This statistic is straightforward to compute, and 
offers a quick test of the consistency of the power spectrum from map to map.

In the approximation that the 2-point function can be treated as a 
multivariate Gaussian distribution, one can form a likelihood function with 
which to estimate the power spectrum normalization.  In \S 2 we present the 
2-point correlation data in the 4-year DMR maps and examine its consistency 
from map to map.  In \S 3 we evaluate the Gaussian likelihood as a function 
of the mean expected quadrupole moment, $Q_{rms-PS}$, under the assumption of 
a scale-invariant, power-law spectrum of anisotropy.  In \S 4 we compare 
these results with those obtained by other methods and summarize our findings.

\section{2-Point Correlation Data}
The 2-point correlation function, as given in Equation \ref{2pt_data_eq}, is 
the average product of all pixel temperatures with a fixed angular separation.
The data are binned into angular separation bins of width $2\ddeg6$ with the 
first bin reserved for all pixel pairs ($i,j$) such that $i=j$, the second bin 
for pairs with separation between $0^{\circ}$ and $2\ddeg6$, and so forth.  
For the present analysis we employ the maps pixelized in galactic coordinates 
and use the custom Galaxy cut described by Bennett et al. (1996), for which 
there are 3881 surviving pixels.  To minimize cosmic variance we assign equal 
weight to each surviving pixel.

\begin{figure}[t]
\psfig{file=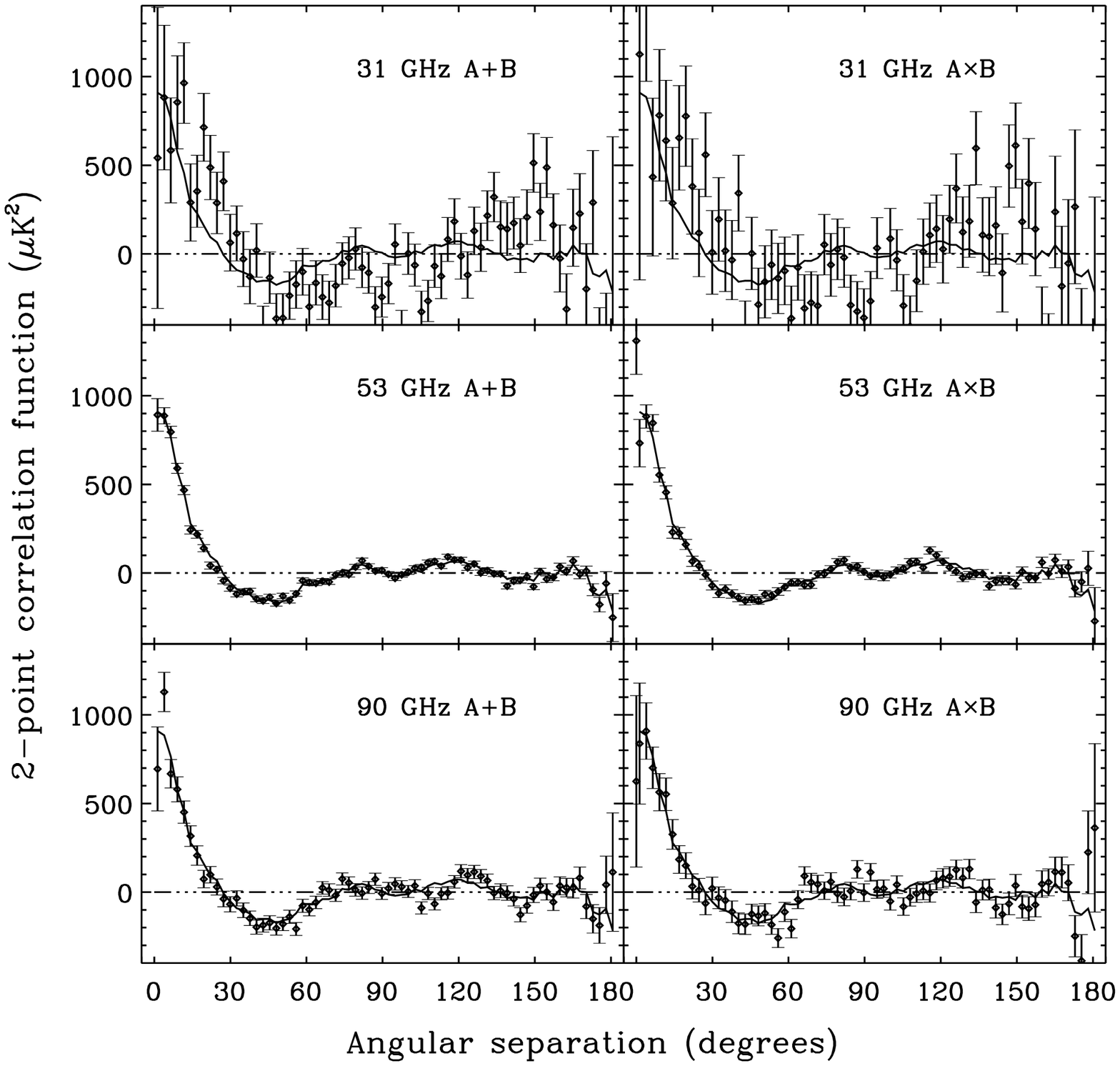,width=6.5in}
\caption{2-point correlation functions obtained from the individual frequency
maps, after monopole and dipole subtraction.  The left-hand panels show the 
auto-correlation functions obtained from a weighted average of the A and B 
channel maps.  The right-hand panels show the cross-correlation of the A and B 
channels, which are sensitive only to common structure in the maps.  The error 
bars represent the uncertainty due to instrument noise, as described in the 
text.  To guide the eye, the solid line is the auto-correlation of the 
weighted average of all six channels maps.  All six 2-point functions are 
statistically consistent with each other.}
\label{315390_fig}
\end{figure}

The basic 2-point functions obtained from the single frequency maps are 
shown in Figure \ref{315390_fig}.  We plot both the auto-correlation of the 
weighted sum of channels A and B at each frequency (the coefficients used to 
form the weighted average maps analyzed in this paper are given in Table 
\ref{map_coeff_table}) and the cross-correlation between channels A and B.
For reference, we also plot, as a solid line, the auto-correlation of the 
weighted average of all six DMR channel maps.  The error bar attached to each 
point represents the $rms$ due to instrument noise, based on 2000 Monte Carlo 
simulations that include only instrument noise.  The plot clearly demonstrates 
excellent consistency of the 2-point correlations at 53 and 90 GHz, even in 
the absence of any galactic signal corrections.  The 31 GHz data exhibit a 
small discrepancy from the mean data that is primarily quadrupolar and is 
presumably due to residual galactic emission.

The data are quantitatively tested for self-consistency by forming differences 
of the 2-point functions and comparing them to simulations.  The statistic for 
the test is defined as 
$\chi^2 = ({\bf\Delta C} - \langle {\bf \Delta C} \rangle)^T \cdot {\bf M^{-1}} 
\cdot ({\bf \Delta C} - \langle {\bf \Delta C} \rangle)$ where ${\bf \Delta C}$
is the observed difference between 2-point functions, with entries 
$\Delta C_a = C^{(1)}(\alpha_a) - C^{(2)}(\alpha_a)$ ($a$ denotes an angular 
separation bin, (1) and (2) denote specific data selections), 
$\langle {\bf \Delta C} \rangle$ is the mean difference, computed from 
simulations described below, and ${\bf M}$ is the covariance matrix computed 
from simulations.  For each realization in the Monte Carlo, we generate a 
single realization of a scale-invariant power-law sky with unit normalization, 
and six noise maps, one per channel, with appropriate noise level and coverage 
(Bennett et al. 1996).  We assume the noise is uncorrelated from pixel to 
pixel, based on the analysis of Lineweaver et al. (1994).  It is then possible 
to generate an ensemble of simulated 2-point functions for any desired auto- or 
cross-correlation function constructable from the DMR data.  We generate such 
an ensemble for each of the six panels depicted in Figure \ref{315390_fig}; 
note that a given realization of the six combinations shares a common CMB 
signal.  We compute $\chi^2$ as defined above for each possible difference 
and compare its value to the ensemble derived from the simulations.  Note that 
our computation of the covariance matrix from simulations automatically 
includes bin-bin correlations in the definition of $\chi^2$.  In no case does 
the observed value of $\chi^2$ exceed the 5\% confidence upper limit derived 
from the simulations, which corroborates the visual consistency of the data.

\begin{figure}[t]
\psfig{file=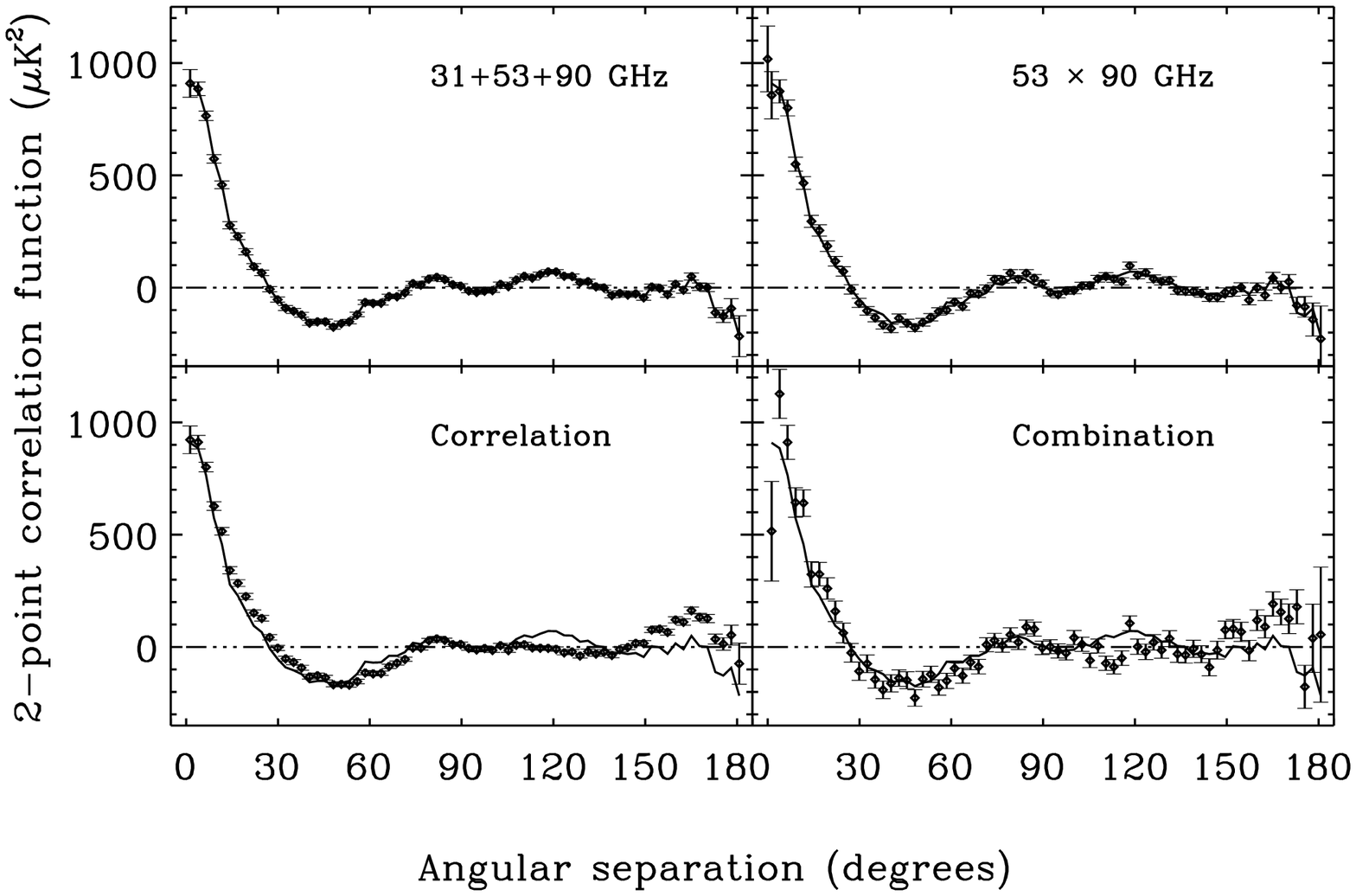,width=6.5in}
\caption{2-point correlation functions obtained from the multi-frequency
maps after monopole and dipole subtraction.  {\it top left}) The 
auto-correlation function of the weighted average map constructed from all 
six DMR channels.  {\it top right}) The cross correlation function of the 53 
GHz weighted sum map with the 90 GHz weighted sum map.  {\it bottom left}) The 
auto-correlation function of the weighted average map with best-fit Galaxy 
template maps subtracted from the map (Kogut et al. 1996a).  {\it bottom right}) 
The auto-correlation function of the linear combination map designed to cancel 
free-free emission.  This map has a best-fit model of the synchrotron and dust 
emission also subtracted.  In all panels the error bars represent the 
uncertainty due to instrument noise, as described in the text.  To guide the 
eye, the solid line is the auto-correlation of the weighted average map 
(top left panel).}
\label{5390ng_fig}
\end{figure}

The 2-point functions obtained from selected multi-frequency combinations of 
the data are shown in Figure \ref{5390ng_fig}.  We plot the auto-correlation 
of the weighted average map, the cross correlation of the 53 and 90 GHz maps, 
and the auto-correlation of two maps which have had residual, high-latitude 
galactic emission modeled and removed (Table \ref{map_coeff_table} and 
Kogut et al. 1996a).  Note the excellent consistency between the 
auto-correlation of the weighted average map, which is sensitive to {\it all} 
structure in that map, and the cross-correlation of the 53 and 90 GHz data, 
which is sensitive only to common structure in the maps.  Note also that the 
two methods used to model and remove high-latitude galactic emission introduce 
only small changes in the 2-point data, and hence in the angular power 
spectrum.  This observation, coupled with the fact that the Correlation and 
Combination model maps render very similar 2-point functions, supports the 
claim of Kogut et al. (1996a) that the free-free emission at high latitudes is 
1) weak, and 2) approximately traced by the DIRBE 140 $\mu$m map at $7^{\circ}$ 
resolution.

\section{Quadrupole Normalization}
Given a power law model of initial Gaussian density fluctuations, 
$P(k)\propto k^n$, where $P(k)$ is the power spectrum of density fluctuations 
as a function of comoving wavenumber $k$, it is possible to derive the 
corresponding angular power spectrum of CMB fluctuations, 
$C_{\ell} = \langle |a_{\ell m}|^2 \rangle$ (Bond \& Efstathiou 1987).  The 
result is
\begin{equation}
C_{\ell} = C_2 \, \frac{\Gamma(\ell+(n-1)/2)\Gamma((9-n)/2)}
                       {\Gamma(\ell+(5-n)/2)\Gamma((3+n)/2)}
\label{qn_model_eq}
\end{equation}
For the scale-invariant case, $n=1$, this reduces to $C_{\ell} = 6 C_2 
/(\ell (\ell + 1))$, which has one free parameter, the mean quadrupole 
moment $C_2$.  We customarily express the normalization in terms of $Q_{rms-PS} 
\equiv \sqrt{(5/4\pi)C_2}$, the mean $rms$ temperature fluctuation expected 
in the quadrupole component of the anisotropy.  We determine the most likely 
quadrupole normalization, $Q_{rms-PS}$, from the 2-point function by evaluating 
the Gaussian approximation to the likelihood function
\begin{equation}
{\cal L}(Q_{rms-PS}) \propto 
\frac {e^{-\frac{1}{2}{\bf \Delta C}^T \cdot {\bf M}^{-1} \cdot {\bf \Delta C}}}
      {\sqrt{\det({\bf M})}}.
\end{equation}
Here ${\bf \Delta C}^T$ and ${\bf \Delta C}$ are $m$-dimensional row and 
column vectors with entries 
$\Delta C_a = C({\alpha}_a) - \langle C({\alpha}_a) \rangle$, and 
${\bf M} = \langle ({\bf \Delta C})({\bf \Delta C})^T \rangle$ is the 
covariance matrix of the correlation function. Note that ${\bf \Delta C}$ is 
defined differently than in \S 2 since we are comparing a single 2-point 
function to an ensemble here.  The angled brackets denote 
averages over both measurement errors and over the ensemble of anisotropy 
fields implied by cosmic variance for a given $C_{\ell}$.  We estimate the 
mean correlation and covariance matrix as a function of $Q_{rms-PS}$ using 
Monte Carlo simulations described above.  The simulations account for 
all important aspects of our data processing including monopole and dipole 
(and quadrupole) removal on the cut sky.  Because of this subtraction, 
the bins of a given correlation function are not all independent so the
covariance matrices derived from the simulations are formally singular.  We 
invert these matrices to form $\chi^2$ using singular value decomposition, 
which permits an unambiguous identification of the zero modes that arise due 
to multipole subtraction.  We then evaluate the logarithm of the likelihood, 
${\ln {\cal L}} = -\frac{1}{2} [\chi^2 + {\ln ({\det({\bf M})})} + const]$, 
in steps of 1 $\mu$K in $Q_{rms-PS}$, spline the result to a resolution of 0.01 
$\mu$K, and identify the maximum.

We test the likelihood method for accuracy by feeding the simulated 2-point 
functions into the likelihood function and solving for an ensemble of 
$Q_{rms-PS}$ maxima.  We define the bias in our method to be 
$\Delta Q = \langle Q_{max} \rangle - Q_{in}$ where $\langle Q_{max} \rangle$ 
is the mean of the recovered maxima and $Q_{in}$ is the simulation input 
normalization.  The resulting bias depends on the noise level in the data, but 
ranges from $-0.2$ to $-0.4$ $\mu$K for the all cases except the 31 GHz data 
where it is $\sim -1$ $\mu$K.  We correct for this bias in all reported 
results.  The uncertainty we assign to $Q_{rms-PS}$ is the $rms$ scatter of 
the ensemble {$Q_{max}$} which typically exceeds the $rms$ of the Gaussian 
likelihood by about 10\%.

The corrected power spectrum normalization deduced from a variety of data 
combinations is given in Table \ref{Q_table}.  The smallest and largest 
normalization deduced from any data combination are 16.4 and 19.6 $\mu$K 
respectively, with values $\sim$18 $\mu$K generally preferred.  The 
normalization inferred from the 2-point function is now in better agreement 
with other determinations than was the case with the 2-year data.  
The change is due to data selection: with the 2-year data, we only analyzed 
the 53 $\times$ 90 GHz cross-correlation function; with the 4-year data we 
have analyzed many more data combinations, including the auto-correlation of a 
weighted average multi-frequency map which yields a normalization $\sim$1.5 
$\mu$K higher than the cross correlation.  The multi-frequency 
auto-correlation is more comparable to the data analyzed by other methods, and 
the 2-point analysis yields consistent results in that case.  For a comparison, 
see Table 2 of Bennett et al. (1996).  In general, the normalization inferred 
from the 4-year data is slightly less than we found after 2 years, in part 
because of the extension of the Galaxy cut.  For comparison, the 31+53+90 GHz 
auto-correlation with a straight 20$^{\circ}$ cut yields a best-fit 
normalization of $Q_{rms-PS}$ = 19 $\mu$K.  As shown in Table \ref{Q_table}, 
the effects of further modeling and subtraction of galactic emission are 
less than 1 $\mu$K in the normalization.

While a likelihood analysis is capable of inferring the best-fit parameters 
for a given model, it does not say anything {\it per se} about the goodness 
of fit.  For reference we have included in Table \ref{Q_table} the values of 
$\chi^2$ at the maximum likelihood value of $Q_{rms-PS}$ (with $n=1$).  Since 
the 2-point function is only approximately multivariate Gaussian distributed, 
our tabulated statistic is only approximately $\chi^2$ distributed.  However, 
we have used our Monte Carlo simulations to compute the expected distribution 
of this statistic and find it to be approximately $\chi^2$ with a mean of 
$\sim70$ and a standard deviation of $\sim12$.  The values computed with the 
DMR data are very consistent with this distribution, implying that the data 
are well fit by a scale-invariant power spectrum.

\section{Conclusions}

Analyses of $Q_{rms-PS\vert n=1}$ have also been reported by Banday et al. 
(1996), G\'orski et al. (1996), Hinshaw et al. (1996), Kogut et al. (1996b), 
and Wright et al. (1996).  All results lie between 15.5 and 19.6 $\mu$K with 
most between 17.5 and 18.5 $\mu$K.  In general, all methods for analyzing a 
given data combination give consistent results, while there is modest 
dependence on data selection.  Fortunately, the dependence on data selection 
does not exceed the statistical uncertainty due to cosmic variance and 
instrument noise.  A reasonable conclusion to draw for the normalization of a 
scale invariant spectrum is $Q_{rms-PS\vert n=1}$ = 18 $\pm$ 1.6 $\mu$K, with 
the error quoted conservatively.

We gratefully acknowledge the many people who made this paper possible: the 
NASA Office of Space Sciences, the {\it COBE} flight operations team, and all 
of those who helped process and analyze the data.

\clearpage
\begin{planotable}{lllllllll}
\tablewidth{6.5in}
\tablecaption{DMR Map Combination Coefficients\tablenotemark{a}}
\tablehead{ \colhead{Map}    &
            \colhead{31A}    &
            \colhead{31B}    &
            \colhead{53A}    &
            \colhead{53B}    &
            \colhead{90A}    &
            \colhead{90B}    &
            \colhead{DIRBE}  &
            \colhead{Haslam} }
\startdata
31ws       & 0.611 & 0.389 & 0     & 0     &  0     & 0     & 0 & 0 \nl
53ws       & 0     & 0     & 0.579 & 0.421 &  0     & 0     & 0 & 0 \nl
90ws       & 0     & 0     & 0     & 0     &  0.382 & 0.618 & 0 & 0 \nl
53+90      & 0     & 0     & 0.412 & 0.299 &  0.110 & 0.179 & 0 & 0 \nl
31+53+90   & 0.049 & 0.032 & 0.378 & 0.275 &  0.102 & 0.164 & 0 & 0 \nl
Correlation\tablenotemark{b} 
       & 0.049 & 0.032 & 0.378 & 0.275 &  0.102 & 0.164 & 3.364 & 0.314 \nl
Combination\tablenotemark{c}
     & $-0.185$ & $-0.117$ & 0.367 & 0.266 &  0.256 & 0.413 & 2.055 & $-0.170$ \nl
\tablenotetext{a}{The maps are formed using the above coefficients according 
to the prescription: $\sum_{i} C_{DMR}^{i} T_{DMR}^{i} - C_H T_H - C_D T_D$, 
where $i = 31A, \cdots, 90B$ is a channel index, $C_{DMR}^i$ are the DMR 
map coefficients given above, $T_{DMR}$ are the DMR maps in $\mu$K of 
thermodynamic temperature, $C_H$ is the coupling coefficient to the Haslam 
map, in $\mu$K / K, given above, $T_H$ is the Haslam map, in K, $C_D$ is the 
coupling coefficient to the DIRBE 140 $\mu$m map, in $\mu$K / (MJy/sr), given 
above, and $T_D$ is the DIRBE 140 $\mu$m map, in MJy/sr.  The resulting map 
has units of $\mu$K, thermodynamic.}
\tablenotetext{b}{The coefficients give the most sensitive combination of the
31, 53 and 90 GHz data, in thermodynamic units.  The Haslam map is used to 
model synchrotron emission, it was fit to the DMR data under the assumption 
that its spectral index was $\beta_s = -3.0$.  The DIRBE 140 $\mu$m map is 
used to model both free-free and dust emission, see Kogut et al. (1996a).  The 
free-free component was fit assuming a spectral index $\beta_{ff} = -2.15$, 
the dust component was fit assuming $\beta_d = +2.0$.  The coefficients given 
above combine the free-free and dust emission.}
\tablenotetext{c}{The coefficients give the most sensitive combination of the
31, 53 and 90 GHz data, in thermodynamic units, consistent with the constraint 
that emission with a spectral index $\beta_{ff} = -2.15$ (free-free emission)
be nullified.  The Haslam map is used to model synchrotron emission, as above.
The DIRBE 140 $\mu$m map is used to model dust emission, as above.}
\label{map_coeff_table}
\end{planotable}

\clearpage
\begin{planotable}{ccllll}
\tablewidth{5.5in}
\tablecaption{Scale-invariant Power Spectrum Normalization}
\tablehead{ \colhead{}                                            &
            \colhead{}                                            &
            \multicolumn{2}{c}{$\ell_{min} = 2$\tablenotemark{b}} &
            \multicolumn{2}{c}{$\ell_{min} = 3$}                  \nl
            \colhead{Map \#1\tablenotemark{a}}                    &
            \colhead{Map \#2\tablenotemark{a}}                    &
            \colhead{$Q_{rms-PS}$ ($\mu$K)}                       &
            \colhead{$\chi^{2}$}                                  &
            \colhead{$Q_{rms-PS}$ ($\mu$K)}                       &
            \colhead{$\chi^{2}$}                                  }
\startdata
\multicolumn{6}{c}{Single frequency cross-correlation}                 \nl
  31A     &    31B    & 18.2 $\pm$ 4.1 & 68.3 & 18.0 $\pm$ 4.6 &  71.0 \nl
  53A     &    53B    & 18.3 $\pm$ 1.6 & 73.5 & 18.6 $\pm$ 1.7 &  69.6 \nl
  90A     &    90B    & 16.4 $\pm$ 2.2 & 72.3 & 18.4 $\pm$ 2.3 &  71.0 \nl
\multicolumn{6}{c}{Single frequency auto-correlation}                  \nl
  31ws    &    31ws   & 17.1 $\pm$ 3.7 & 67.9 & 17.6 $\pm$ 4.0 &  79.6 \nl
  53ws    &    53ws   & 18.7 $\pm$ 1.6 & 99.9 & 19.4 $\pm$ 1.6 &  97.5 \nl
  90ws    &    90ws   & 17.5 $\pm$ 2.0 & 63.5 & 19.0 $\pm$ 2.2 &  61.5 \nl
\multicolumn{6}{c}{Multi-frequency cross-correlation}                  \nl
  53ws    &    90ws   & 17.2 $\pm$ 1.5 & 60.8 & 17.8 $\pm$ 1.5 &  64.1 \nl
  53ss    &    90ss   & 17.0 $\pm$ 1.6 & 61.2 & 17.9 $\pm$ 1.6 &  62.3 \nl
\multicolumn{6}{c}{Multi-frequency auto-correlation}                   \nl
53+90     & 53+90     & 18.5 $\pm$ 1.4 & 84.2 & 19.6 $\pm$ 1.5 &  83.5 \nl
31+53+90  & 31+53+90  & 18.6 $\pm$ 1.4 & 80.0 & 19.3 $\pm$ 1.4 &  78.2 \nl
\multicolumn{6}{c}{Multi-frequency auto-correlation with Galaxy model} \nl
Correlation & Correlation 
                      & 17.5 $\pm$ 1.4 & 76.2 & 18.5 $\pm$ 1.4 &  78.3 \nl
Combination & Combination 
                      & 16.7 $\pm$ 2.0 & 89.2 & 17.8 $\pm$ 2.2 &  92.4 \nl
\tablenotetext{a}{The coefficients that comprise the map combinations in these
columns are given in Table \ref{map_coeff_table}, except for 53ss and 90ss 
which are straight sum maps: (A+B)/2.}
\tablenotetext{b}{$\ell_{min}$ is the lowest order multipole remaining in the 
map after subtracting a best-fit multipole of order $\ell_{min}-1$.  
$Q_{rms-PS}$ is the most-likely quadrupole normalization, after calibrating 
the likelihood with Monte Carlo simulations.  $\chi^2$ is tabulated, for 
reference, with respect to the mean of a scale-invariant model with the 
corresponding most-likely normalization.  There are 71 bins in the 2-point 
function.}
\label{Q_table}
\end{planotable}

\clearpage

\clearpage
\begin{references}
\reference Banday, A.J., et al. 1996, \apjl, in preparation
\reference Bennett, C.L., et al. 1992, \apjl, 396, L7
\reference Bennett, C.L., et al. 1994, \apj, 436, 423
\reference Bennett, C.L., et al. 1996, \apjl, submitted
\reference Bond, J.R., \& Efstathiou, G. 1987, \mnras, 226, 655
\reference G\'orski, K.M., Hinshaw, G., Banday, A.J., Bennett, C.L., 
 Wright, E.L., Kogut, A., Smoot, G.F., \& Lubin, P. 1994, \apjl, 430, L89
\reference G\'orski, K.M., Banday, A.J., Bennett, C.L., Hinshaw, G., Kogut, 
 A., Smoot, G.F., \& Wright, E.L. 1996, \apjl, submitted
\reference Hinshaw, G., Banday, A.J., Bennett, C.L., G\'orski, K.M., Kogut, 
 A., Smoot, G.F., \& Wright, E.L. 1996, \apjl, submitted
\reference Kogut, A., et al. 1992, \apj, 401, 1
\reference Kogut, A., Hinshaw, G., Banday, A.J., Bennett, C.L., G\'orski, K.M., 
 Smoot, G.F., \& Wright, E.L. 1996a, \apj, submitted
\reference Kogut, A., Banday, A.J., Bennett, C.L., G\'orski, K.M., Hinshaw, G., 
 Smoot, G.F., \& Wright, E.L. 1996b, \apj, submitted
\reference Lineweaver, C.H., et al. 1994, \apj, 436, 452
\reference Smoot, G.F., et al. 1992, \apjl, 396, L1
\reference Wright, E.L., et al. 1992, \apjl, 396, L13
\reference Wright, E.L., Smoot, G.F., Bennett, C.L., \& Lubin, P.M. 1994,
 \apj, 436, 443
\reference Wright, E.L., Bennett, C.L., G\'orski, K.M., Hinshaw, G., \&
 Smoot, G.F. 1996, \apjl, submitted
\end{references}
\end{document}